**Title**

**Study on Locomotive Epidemic Dynamics in a Stochastic Spatio-Temporal Simulation Model on a Multiplex Network**

**Authors**

H.M. Shadman Tabib[1*], Jaber Ahmed Deedar[1*], K.M. Ariful Kabir[2]

[1]Department of Computer Science and Engineering, Bangladesh University of Engineering and Technology, Dhaka-1000, Bangladesh.

[2]Department of Mathematics, Bangladesh University of Engineering and Technology, Dhaka-1000, Bangladesh.

**\* equal contribution**

**Corresponding author**

H.M. Shadman Tabib,shadmantabib2002@gmail.com,2005103@ugrad.cse.buet.ac.bd

**Abstract**

This study presents an integrated approach to understanding epidemic dynamics through a stochastic spatio-temporal simulation model on a multiplex network, blending physical and informational layers. The physical layer maps the geographic movement of individuals, while the information layer tracks the spread of knowledge and health behavior via social interactions. We explore the interplay between physical mobility, information flow, and epidemic outcomes by simulating disease spread within this dual-structured network. Our model employs stochastic elements to mirror human behavior, mobility, and information dissemination uncertainties. Through simulations, we assess the impact of network structure, mobility patterns, and information spread speed on epidemic dynamics. The findings highlight the crucial role of effective communication in curbing disease transmission, even in highly mobile societies. Additionally, our agent-based simulation allows for real-time scenario analysis through a user interface, offering insights into leveraging physical and informational networks for epidemic control. This research sheds light on designing strategic interventions in complex social systems to manage disease outbreaks.

**Introduction:**

Recently, the world has witnessed the devastating effects of epidemic outbreaks, such as the COVID-19 pandemic [1]. Understanding and modeling the spread of infectious diseases is crucial for implementing effective control measures and mitigating their impact on public health [2]. Traditional epidemic models, such as the classic susceptible-infected-recovered (SIR) model [3], provide valuable insights into disease dynamics. However, these models often overlook the influence of individual movement patterns on disease transmission. To address this limitation, we propose a novel SIR epidemic model that incorporates individual mobility between two distinct locations: home and office. By considering the movement of individuals, we aim to capture the realistic dynamics of disease transmission in a more accurate and detailed manner. This model allows us to explore the interplay between individual mobility [4], contact patterns, and the spread of infection, thereby enabling us to devise targeted disease control and prevention strategies . The central idea of our model is that individuals move between their home and office, interacting with others in both locations. The infection probability depends on infected individuals' presence in their immediate surroundings or "neighbors." This approach acknowledges that disease transmission is more likely to occur in proximity and can be influenced by contact rates within home and office settings. By accounting for spatial and social factors [5], we can gain deeper insights into disease spread mechanisms.

According to the report available on the World Health Organization (WHO) website [6], globally, after the continued decline observed since the end of March 2022, new weekly COVID-19 cases have stabilized during the reporting period (9 May to 15 May 2022), with over 3.6 million cases reported, a 1% increase as compared to the previous week. The number of new weekly deaths continues to decline, with over 9000 new deaths reported during the same period, representing a 21% decrease as compared to the previous week. In response to the ongoing challenge of modeling epidemic outbreaks, the susceptible-infected-recovered (SIR) model has proven to be a reliable tool for understanding disease dynamics. Initially introduced by McKendrick and Kermack [7], the SIR model has evolved, incorporating various enhancements and additions to improve its accuracy and effectiveness. One such enhancement is the utilization of agent-based models [8], where individual agents possess specific parameters that influence the overall simulation of the system. Previous works have explored SIR models using static networks, such as ring networks [9][10], Erdős-Rényi random networks [11], and random regular networks [12][13][14]. These models have been analyzed using the Gillespie algorithm [15][16][17], a widely used stochastic simulation algorithm. Additionally, there have been significant advancements in multiplex networks, where agents are categorized into different layers (e.g., physical and information layers), and their interactions influence the system dynamics over time using Markov chain processes [18][19][20]. Beyond COVID-19, stochastic models have been developed for various other diseases, including HIV, Ebola, Plague, and Cholera, utilizing mathematical simulations and equations [21][22][23][24]. Some studies have also focused on leveraging real-life data to predict and optimize the implementation of lockdown measures in specific areas [25].

Agent-Based Modeling (ABM) is a powerful technique for simulating complex systems through interactions between agents and their environment [26]. Agents and the environment have parameters and states that can change over time based on defined rules. This approach enables data collection and the visualization of various parameters over time, often revealing unexpected emergent behaviors. ABM demonstrates how simple rules can lead to complex outcomes. Our ABM implements a stochastic SIR model with adjustable parameters, allowing us to analyze the impact of different factors on epidemic dynamics. The model provides real-time information on susceptible, infected, and recovered agents, allowing for the graphical representation of these dynamics over time.

Our model extends beyond interpersonal contacts and encompasses various critical factors influencing infection dynamics. Firstly, it accounts for individuals' ability to stay at home, reducing their mobility and potential virus exposure, which can be influenced by public health guidance, government policies, or personal choices. Analyzing this behavior helps assess containment measures' effectiveness. Secondly, we incorporate population density, directly impacting the likelihood of person-to-person encounters. High-density areas foster more interactions and, consequently, increase disease transmission risk. By considering population density, we evaluate urbanization's role in infection spread. Lastly, we include movement spread, which measures how infected individuals can transmit the virus during travel, potentially leading to new outbreaks in unaffected areas. Accounting for movement spread helps us understand the role of transportation networks [27] and long-distance travel in epidemic dynamics [28]. Our model is based on established knowledge from epidemiology, mathematical modeling [29], and network science [30]. It builds upon prior research on individual movement in epidemic modeling, like commuting patterns, transportation networks, and social contact networks [31].

Our model involves mobile agents with defined collision radii, and random collision probabilities govern their interactions during the simulation. Agent collisions represent meaningful interactions between them. We monitor the counts of susceptible, infected, and recovered individuals across multiple adjustable parameters, which are compared to the traditional SIR model [32]. In our model, mobile agents move within the space, temporarily residing in specific grid cells. The infection rate of these cells increases as more infected agents occupy them and decreases with the virus's half-life [33]. We have implemented an interactive user interface for our agent-based model, allowing users to customize parameters and generate real-time graphs during each simulation. This approach offers a more user-friendly and efficient way to approximate epidemic dynamics in the real world while introducing enhancements to existing models.

In summary, our novel SIR epidemic model considers individual mobility between home and office settings. By incorporating movement, contact patterns, staying at home, population density [34], and the movement spread, it provides a comprehensive framework for studying disease transmission. This research enhances our understanding of real-world disease dynamics, informs public health interventions, optimizes resource allocation, and mitigates the impact of future epidemics.

**Model and Method**

In the traditional SIR model, the total number of susceptible, infected, and recovered individuals remains constant, with transitions represented using deterministic ordinary differential equations (ODEs) [35]. In our approach, we implement an enhanced stochastic SIR model simulation. Agents are analogous to randomly moving objects in a defined space, with their collision probabilities being random. These "agents" are analogous to "susceptible," "infected," and "recovered" entities.

Initially, these agents are distributed randomly in space, and their interactions create a network of nodes and edges. When a susceptible agent transitions to an infected state, an edge is formed between them. This network can be visualized in various arrangements, such as ring, random, spring, and planar graphs, showcasing the maximum probability of these randomly distributed vertices. Additionally, in the information layer, agents are connected to a maximum number of friends in a "small world" network scheme [36]. The random movements of these agents lead to collisions, which exhibit multiple features.

Figure 2 represents three panel comprising three phases which display the visual representations and flowcharts associated with the model. This model consists of two principal components, namely agents and the environment. Both agents and the environment possess specific parameters and states. The user can adjust parameters before the simulation and remain constant throughout the simulation, while states undergo changes during the simulation and are not user adjustable. Within this model, agents within the environment generate information and a physical layer simultaneously, and the interaction between agents and the environment collectively constitutes a spatial layer. The simulation process unfolds in four distinct phases: Initialization (where agents and the environment are set up), the main loop encompasses the simulation steps, agent-environment interaction (representing passive infection originating from the environment), and agent-agent interaction, depicting the direct transmission of infections among agents.

In this simulation, the environment is initialized as a 2D grid with fixed dimensions, serving as the space for agent movement. Initially, all cells have zero infection rates, which increase when infected agents visit them. The environment's parameters include agent density, the proportion of motionless agents, and the virus's half-life. The color of a cell represents its infection rate, with lighter colors indicating higher infectiousness.

Agents are also initialized with specific parameters. They are represented as circles with a fixed radius. When two agents intersect, it's considered an interaction that can lead to infection of susceptible agents, creating a physical connection between them. The radius of an agent's circle affects its infection rate. Agents can also get infected from the environment. Each agent has a home and office cell, initially positioned at home. Agents move back and forth along a straight path between these locations. Agents have various states, including infection status, friends, days infected, and awareness, indicated by their color.

The main loop of the simulation involves several steps. First, the infection rates of cells decay over time based on a half-life formula. For each agent, their interactions with the environment are simulated. Additionally, interactions between distinct pairs of agents are checked, optimizing performance using a grid. Agent-environment interactions include randomly selecting a friend, and if that friend is aware, the current agent becomes aware with a certain probability, reducing their infection risk. If an agent is recovered, their status remains unchanged, and if they are infected, their cell's infection rate increases, contributing to disease spread in the environment. Agents recover if their infection period exceeds their recovery time. Uninfected agents can become infected with a probability related to their cell's infection rate, representing environmental transmission. Agents move between their home and office cells at a specified velocity.

In the context of the simulation, agent-agent interactions are processes where we assess how two agents, labeled agent A and agent B, interact with each other. Specifically, we are interested in determining if Agent A is close to Agent B within a certain distance, known as the infection radius. There are two scenarios to consider during these interactions. If Agent A is already infected or has previously recovered from an infection, no further action is taken. In this case, agent A is considered immune to further infections, and the interaction doesn't lead to any changes in their infection status. However, if Agent A is within the infection radius of another agent, Agent B, who is currently infected, a different outcome occurs. In this situation, Agent A becomes infected. This means agent A contracts the infection from the nearby infected agent B. This process mimics the direct transmission of a disease from one agent to another, and agent A's infection status changes from susceptible to infected.

**Results and Discussions**

We conducted a detailed numerical investigation into the Local Epidemic Dynamics (LED) model, presenting our findings through line graphs. These graphs illustrate the behavior of our innovative epidemic model as we vary its parameters. The effectiveness of epidemic control policies depends on these parameters and their respective frequencies. To understand this relationship better, we examined line graphs depicting the number of infected individuals under two scenarios: one without any information dissemination and the other with an information layer incorporated into the control measures. This allowed us to compare the outcomes of the epidemic under different conditions. Additionally, we plotted the Final Epidemic Size (FES) corresponding to each parameter for every case studied. This approach provides a direct correlation between each parameter and the overall extent of the epidemic. By doing so, we gain insights into how changes in these parameters can influence the severity and spread of the epidemic.

Our numerical analysis investigated the impact of epidemic dynamics for a simple SIR model by deactivating the information layer. Figure 3 offers a series of simulation snapshots that depict the temporal dynamics of susceptible (blue), infected (red), and recovered (green) agents. As the simulation unfolds, it becomes evident that the pandemic gradually recedes, culminating in a scenario where only recovered individuals remain. A defining characteristic of our model is the mobility patterns exhibited by the agents, specifically, their transitions between home and office settings, as illustrated in Figure 1. During these movements, agents traverse within their contact radius, leading to potential infections through direct physical interactions. A range of parameters, including velocity, recovery time, infection rate, the initial number of infected individuals, virus half-life, and population density, influence the epidemic's propagation. Notably, the information layer has yet to be excluded in this particular analysis. The variability in these parameters yields significant differences in the total number of infected individuals and the final size of the epidemic. The information and spatial layers remain inactive throughout the simulation snapshots presented. Agents' states—susceptible, infected, and recovered—are denoted by blue, red, and green colors, respectively. Initially, the predominant blue color signifies a high number of susceptible individuals. Subsequently, the red color begins to increase, indicating an increasing number of infections, while the emergence of green suggests that some infected individuals are recovering. Ultimately, most of the population transitions to the recovered state, leading to a predominance of green coloration in the simulation.

This study compares infection counts across three epidemic models for different population sizes: (i) $N = 100$, (ii) $N = 1000$, (iii) $N = 5000$, and $N = 10000$. Figure 4 clearly illustrates the distinctions in total infection numbers among the three models: the ODE-based SIR model [37], the Gillespie Algorithm-based SIR model, and our proposed Locomotive Epidemic Dynamics (LED) model. Notably, the infection curves generated by the Gillespie and LED methods closely overlap. For all simulations using the LED model, parameters were set to their standard values: an infection rate of 0.5, recovery time of 10 days, velocity of 5 pixels per second, density set to 1 (indicating agents form a complete graph), no stationary agents (all agents were mobile), a data rate of 0.5 (reflecting the probability of an individual becoming aware after interacting with an informed individual), a viral half-life of 0.0000001 days, a maximum of 10 friends per agent, with one initially aware agent and three initially infected agents. Thus, our comparison of infection dynamics across three epidemic models—ODE-based SIR, Gillespie Algorithm-based SIR, and the Locomotive Epidemic Dynamics (LED) model—demonstrates apparent differences in infection patterns, with the Gillespie and LED models showing highly similar results. With its parameter set reflecting realistic infection and recovery rates, agent movement, and awareness dynamics, the LED model offers a robust framework for simulating the spread of infectious diseases in dynamic populations. This model highlights the importance of incorporating agent mobility and awareness in understanding epidemic outcomes.

Figure 5 illustrates the relationship between the spread of an epidemic and key parameters in the absence of an information layer. In figure 5(i), as agent velocity increases, the likelihood of collisions between moving agents also rises, leading to a larger epidemic size at higher velocities. This suggests that faster-moving agents facilitate more significant interaction, resulting in more disease transmission. Figure 5(ii) shows the impact of recovery time on the epidemic. A longer recovery means that infected individuals remain contagious for a more extended period, increasing the number of infections and, consequently, the final epidemic size (FES). This highlights the importance of shorter recovery periods in controlling the spread of disease, as quicker recoveries can help reduce the overall infection rate. Finally, in figure 5(iii), the effect of infection rate on epidemic spread is evident. A higher infection rate leads to more infected individuals, and the FES increases accordingly. This is because a higher infection rate raises the probability of disease transmission, significantly amplifying the spread of the epidemic. The analysis shows that higher agent velocity, longer recovery times, and higher infection rates contribute to larger epidemic sizes, emphasizing the need for control measures to mitigate these factors in disease management.

Figure 6 also presents key insights into how different initial conditions and parameters influence the spread of an epidemic without an information layer. Figure 6(i) shows that a more significant number of initially infected individuals at the beginning of an outbreak significantly increases the likelihood of direct and indirect transmission. As a result, the final epidemic size (FES) grows with a higher initial infection count. This highlights the critical role of early containment efforts, as more infected individuals at the start allow the epidemic to spread more rapidly and broadly. Figure 6(ii) explores the impact of virus half-life, particularly in indirect transmission via environmental contamination, referred to as the "space layer." As the half-life of the virus increases, so does the FES. This indicates that the virus remains viable in the environment for more extended periods, allowing more opportunities for susceptible agents to become infected by entering contaminated areas. For instance, if an infected individual sneezes in a particular space, the virus persists for a duration governed by its half-life. Any agent moving through this contaminated space before the virus becomes inactive will likely contract the infection. This transmission layer adds a spatial dimension to the dynamics of disease spread in the model. Figure 6(iii) examines the effect of agent density on epidemic progression. In this simulation, agents are represented as circles, and the density is defined using the equation:

$$\text{Density} = \frac{\text{Total frame Area}}{\pi r^2}$$

Where $r$ is the contact radius, representing the distance between which agents are in contact with one another, as the density of agents increases, the number of infected individuals also rises. Higher density implies more frequent interactions and closer proximity between agents, increasing the chances of transmission. The graphs in figure 6(iii) demonstrate that for each increase in density, the number of infected agents grows, leading to a corresponding rise in FES. Therefore, the results depicted in figure 6 emphasize the significant influence of initial infection levels, virus half-life, and agent density on epidemic outcomes. Higher initial infections, longer virus viability in the environment, and increased population density all contribute to a larger epidemic size, underscoring the importance of controlling these factors in epidemic management.

We now extend our analysis of infectious disease dynamics by incorporating an active information layer, a factor that absents in previous discussions. Figure 7 summarizes the key results of integrating this information layer into the model. We examine the variation in Susceptible, Infected, and Recovered populations under conditions where agents are mobile, and the information layer is active. Panel (a) illustrates a scenario where agents move between home and work while maintaining virtual connections. These virtual ties, depicted by yellow lines, represent friendships that enable communication about the infectious disease. Aware agents, represented by blue circles, have a reduced contact radius, lowering their probability of infection compared to unaware agents, shown by red circles with a larger contact radius. Panel (b) visualizes the spread of the disease through color-coded agents: blue for Susceptible, red for Infected, and green for Recovered individuals. This graph captures disease progression over time, with the gray-shaded areas representing the environmental 'Space' layer, which facilitates indirect transmission. This spatial layer increases the likelihood of infection among susceptible individuals, a phenomenon analyzed in detail in later sections.

Panel (c) presents three comparative graphs. The first shows the overall dynamics of Susceptible, Infected, and Recovered populations over time, incorporating both aware and unaware agents. The second and third graphs directly compare infection dynamics with and without the active information layer, focusing on the Susceptible and Infected populations. The results reveal a significant distinction in the behavior of aware versus unaware agents, affirming our assumption that awareness reduces the probability of infection. The simulation offers a more profound perspective by integrating the information and spatial layers. Aware susceptible agents are depicted in yellow, while aware infected individuals are shown in cyan, each with a reduced contact radius. Additionally, white cells represent areas where the virus remains active due to its persistence in the environment after an infected individual has been present. The simulation captures the epidemic's progression over time, with the initial prevalence of blue (susceptible population) gradually giving way to red (infected) and green (recovered) as the outbreak unfolds. Toward the conclusion, the dominance of green indicates that a large proportion of the population is recovering. At the same time, the presence of yellow and cyan dots with reduced radii highlights the impact of awareness in limiting the spread of infection.

In the subsequent analysis, we examine the effects of various parameters on disease dynamics, as illustrated in figures 8 and 9. The first graph (i-a) in figure 8 shows the number of infected individuals over time as the velocity parameter varies, with the information layer active. A higher velocity increases the probability of collision between agents, leading to a higher infection rate. This is evident from the steep curve observed when the velocity is set to 50. As the speed of the velocity decreases, the number of infections drops significantly. At a velocity 1 (one), the infection curve nearly reaches zero, indicating minimal spread. The second graph (i-b) in figure 8 depicts the Final Epidemic Size (FES) about velocity. When the information layer is active, a slight decrease in FES is observed at lower velocities. However, as velocity increases, the effectiveness of the information layer diminishes. Thus, we conclude that the information layer reduces infection only when the agents' velocity is low.

Figure 8 (ii-a) shows the number of infected agents over time while varying the recovery time parameter with the information layer active. A very short recovery time leads to a sharp reduction in infected individuals, almost zero. Conversely, increasing recovery time results in a higher infection rate. The second graph (ii-b) examines FES as a function of the recovery time. Here, the information layer only significantly impacts FES when recovery time is varied. In figure 8(iii-a), higher infection rates lead to a steeper increase in infected individuals over time. The graph in figure 8(iii-b) compares the FES with and without the information layer, demonstrating that the presence of an active information layer results in a lower FES, underscoring its role in reducing infections.

In figure 9(i-a), we analyze the effect of the initial number of infected agents on the spread of the disease. A higher initial number of infected agents leads to a more rapid spread of the infection, as they can infect more susceptible individuals. In figure 9(i-b), we compare the FES with and without the information layer, which reveals that the FES is lower when the information layer is active, indicating the effectiveness of awareness in controlling the epidemic. The half-life of the virus, which contributes to the environmental transmission in the spatial layer, is analyzed in figure 9(ii). The FES graph in figure 9(ii-b) shows a reduced tendency for infection spread when the information layer is active compared to its absence. Figure 9(iii-a) presents the number of infected individuals over time, showing that as population density increases, the infection curve becomes steeper, resulting in a faster rise. The second graph (iii-b) compares the FES for different population densities. It demonstrates that the active information layer consistently reduces FES, highlighting its crucial role in mitigating infections when agents share information virtually. Thus, the analyses show how varying parameters—such as agent velocity, recovery time, infection rate, initial infected population, virus half-life, and population density—impact the overall infection dynamics over time. The simulations were conducted using a population of 10,000 agents, with each result averaging over ten simulation runs to ensure statistical reliability and represent the most probable trends for each parameter.

This framework, coupled with a user-friendly interface that supports parameter customization (as described in the Supplementary Section), is an effective instrument for simulating and examining how agent mobility and different layers of information impact the propagation of diseases. The UI includes options for visualizing movement and information network graphs, enabling users to experiment with different configurations of nodes and networks. This flexibility allows predictions and strategic interventions in real-world epidemics by evaluating various parameters specific to a given population and locality. The model and interactive interface offer a potential solution for understanding and managing complex interactions among agents, considering both mobility and virtual information networks simultaneously.

**Conclusion**

This study presents a novel Locomotive Epidemic Dynamics (LED) model, which couples physical and information layers to provide a more comprehensive understanding of epidemic dynamics in complex social systems. These two layers are interlinked through various real-world parameters, reflecting the interaction between individual mobility and information dissemination during an outbreak. The LED model introduces significant advancements over traditional simulation approaches, such as the baseline SIR model, by incorporating agent mobility and an information layer alongside the conventional epidemic dynamics.

In our analysis, we systematically explored the impact of the information layer by first deactivating it and varying key parameters to examine disease spread solely based on physical movement. This was followed by activating the information layer, and we observed notable differences in the epidemic outcomes. The comparison between these two scenarios—information layer off versus information layer on—was quantified by scaling the "Infected" axis between 0 and 1 to facilitate direct comparison with existing models. Furthermore, we examined the Final Epidemic Size (FES) under both conditions, demonstrating that fewer individuals become infected when the information layer is active throughout the epidemic.

The results reveal that the active information layer is critical in reducing the number of infections by enhancing awareness, influencing behavior, and promoting timely preventive actions. This is reflected in the significant reduction in the FES when the information layer is integrated, underscoring the importance of communication and information flow in managing epidemics. Our simulations with realistic parameter ranges show that incorporating information dissemination with physical mobility leads to more predictable and controlled epidemic outcomes.

The LED model offers valuable insights into how epidemic spread can be mitigated by leveraging both the physical movement of individuals and the strategic dissemination of information. Our approach provides a powerful tool to predict contagion propagation paths and diffusion patterns in real-world scenarios using graph-based models to represent these two layers. The findings highlight the importance of adopting context-specific preventive measures—such as lockdowns, social distancing, and communication campaigns—considering the complex interactions between agents, mobility patterns, and information flow.

Overall, our model contributes a new dimension to epidemic modeling by integrating real-world parameters and emphasizing the active role of the information layer in controlling disease spread. The combination of agent-based mobility and information dissemination provides a framework for designing effective intervention strategies tailored to specific localities. Policymakers and health authorities can use these insights to implement more efficient and targeted measures, ultimately enhancing epidemic control efforts in complex and dynamic environments.

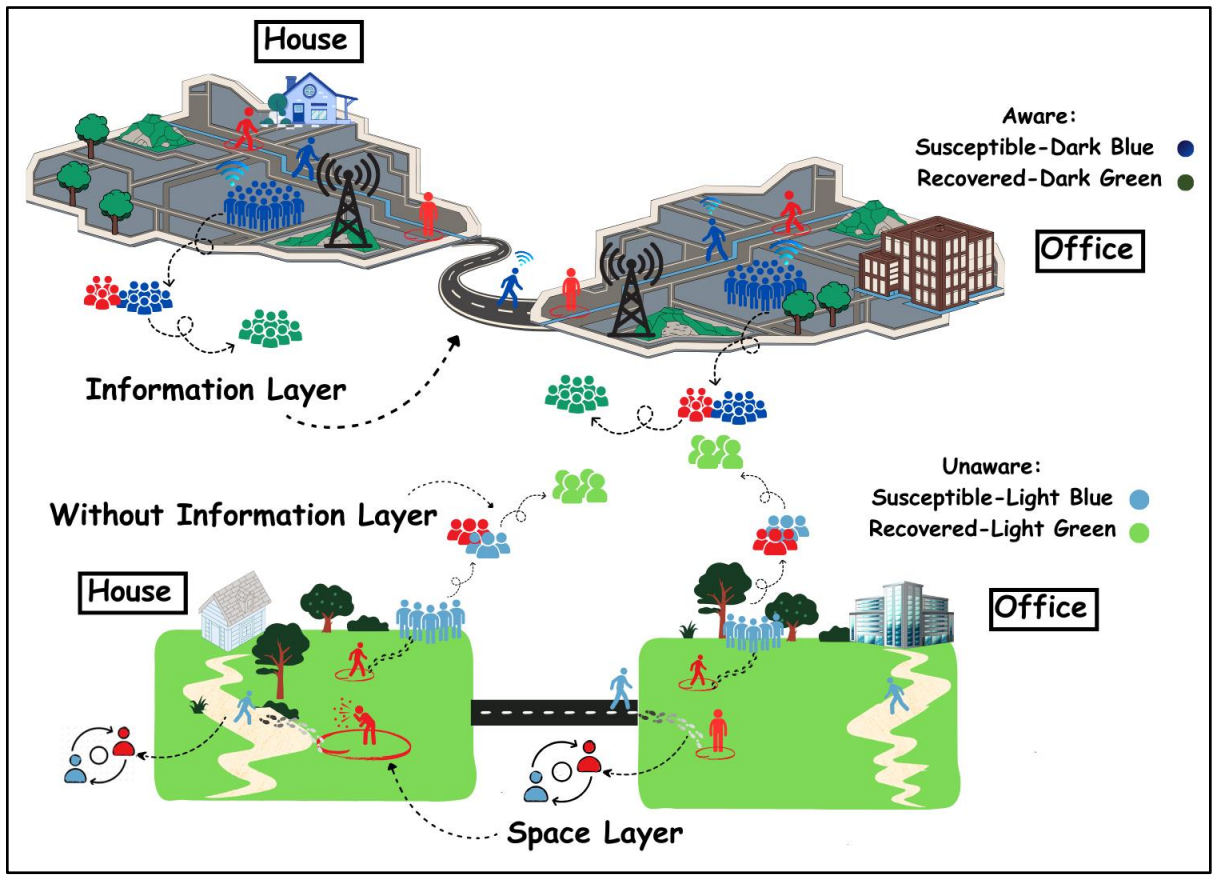

**Figure 1.** This model compares scenarios with active and inactive information layers to represent the propagation of an epidemic. Without an information layer, agents are free to travel around and spread infection when they come into contact with sick people. This results in an increase in the number of infections. Susceptible agents become aware of the epidemic, take preventative measures, and limit close interactions when the information layer is active, which lowers infection rates. The model accurately simulates real-world epidemic dynamics by contrasting these two scenarios visually and showing how informed agents, who modify their behavior, experience slower disease transmission and fewer overall infections compared to unaware agents.

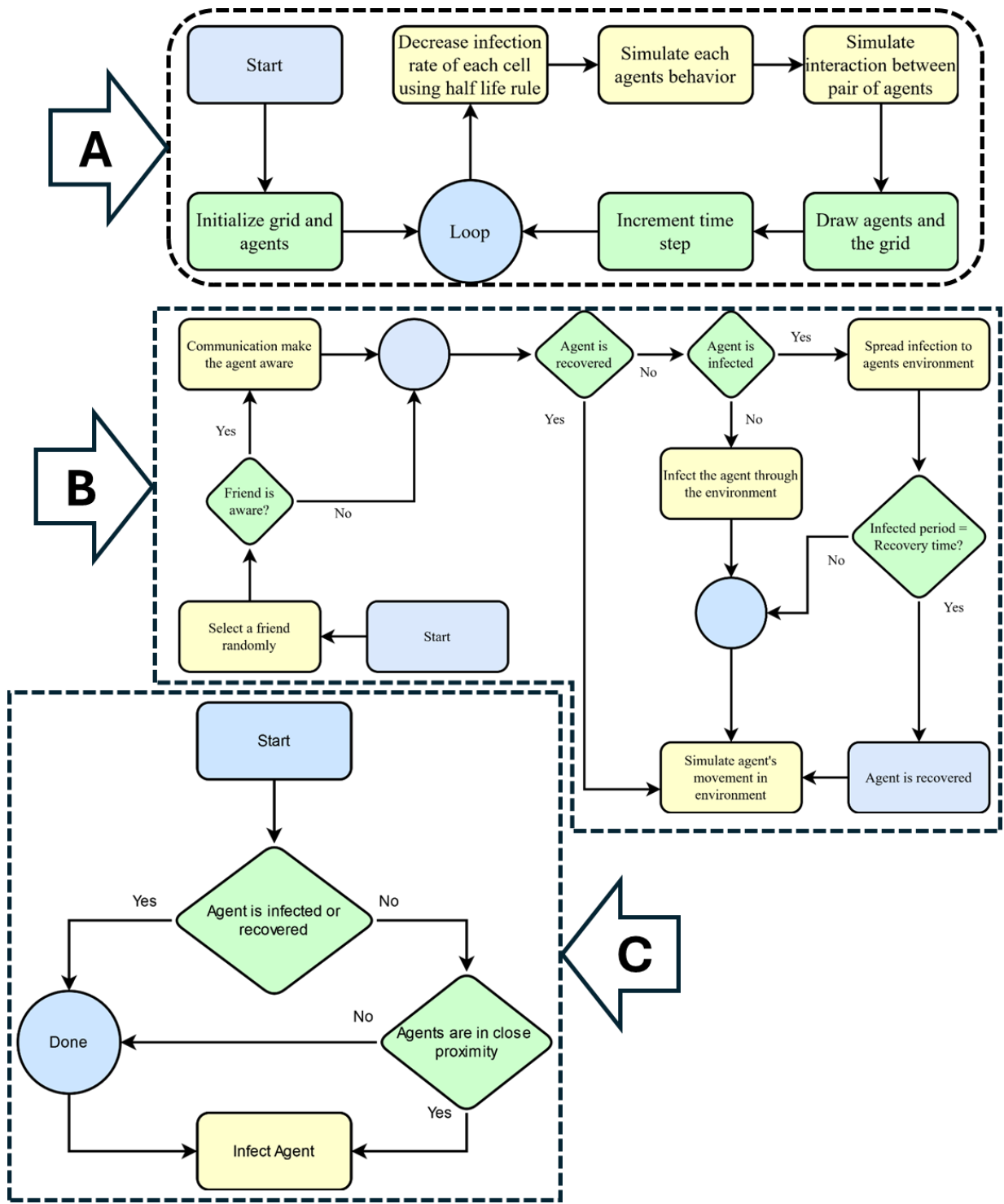

**Figure 2. The workflow of the simulations. Panel A. Main Flow Process**. This panel illustrates the primary flow of the simulation, starting with agents (classified as susceptible, infected, or recovered) moving and interacting within a 2D grid. At each loop step, the half-life rule is applied, reducing the infection rates as agents move between their homes and offices, communicate with one another, and potentially contract the infection from other agents or the environment. If a susceptible agent comes into contact with an infected agent during these interactions, there is a chance of transmission. The simulation continues iteratively, updating the grid and agents' states until it concludes or there are no remaining infections. **Panel B. Agent Behavior Simulation.** This section demonstrates the simulation of an agent's behavior. The process starts by randomly selecting a friend. Communication alerts the agent if this friend is aware of the infection, decreasing their risk of contracting the disease. If the agent has recovered, they are immune; otherwise, they risk infecting those around them. The environment may infect the agent even if no other agents are involved. The agent recovers if the infection and recovery periods match. This ongoing cycle of movement and interaction captures the system's dynamic state changes and behaviors. **Panel C. Agent-Interaction Process.** This panel depicts the process of interactions between agents—first, the simulation checks if an agent is infected or has recovered. If either condition is met, no further actions are taken in that iteration. If the agent is susceptible, the simulation then checks if the agents are close enough to interact. If not, the interaction terminates. However, if the agents are nearby, the susceptible agent becomes infected, and the interaction is concluded. This process captures the disease transmission dynamics based on the spatial proximity of agents.

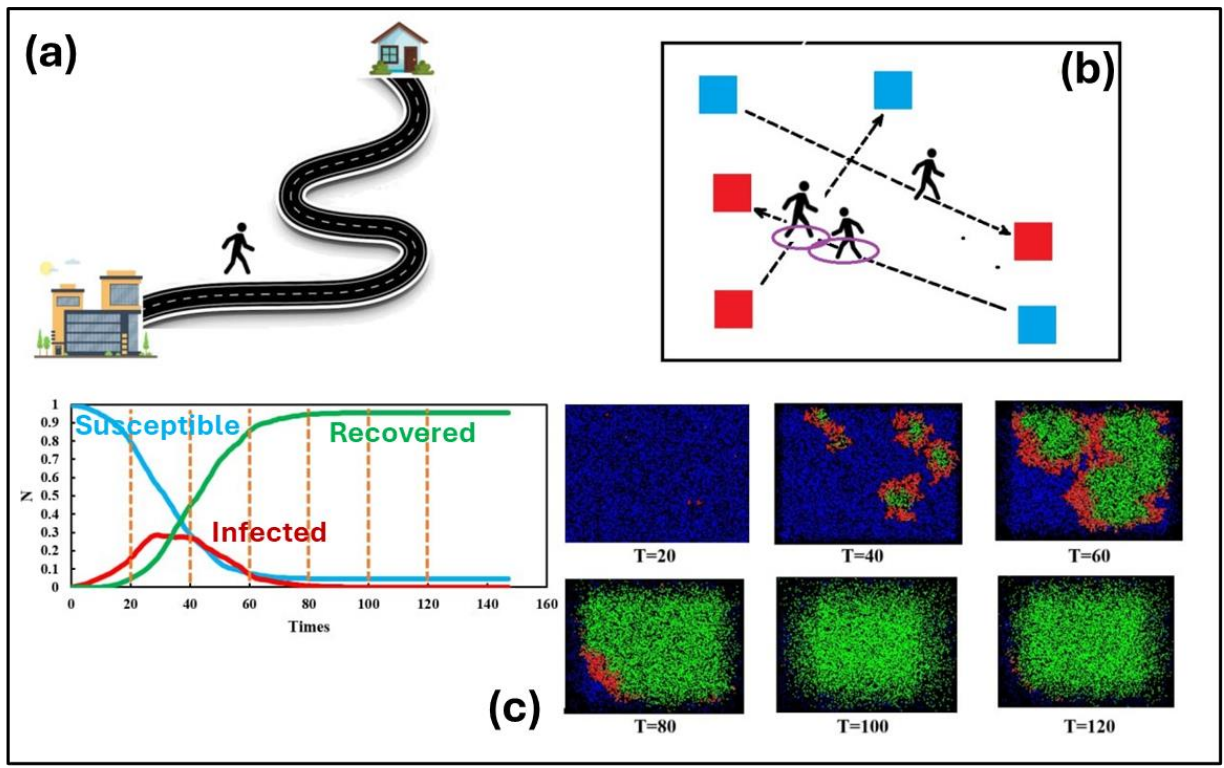

**Figure 3.** The SIR model is present without an information layer, focusing on agent movement. In panel (a), agents move between home and office without any virtual connections or awareness of others. Panel (b) shows how agents come into physical proximity during movement, forming a contact network based on the intersection of their contact information, representing potential disease transmission points. In panel (c), the time evolution of the epidemic is displayed, with blue, green, and red circles representing susceptible, recovered, and infectious individuals, respectively. These circles depict the agent populations at various time intervals (t = 20, 40, 60, 80, 100, and 120), offering a straightforward visual narrative of the epidemic's initiation, growth, peak, and eventual decline as time progresses.

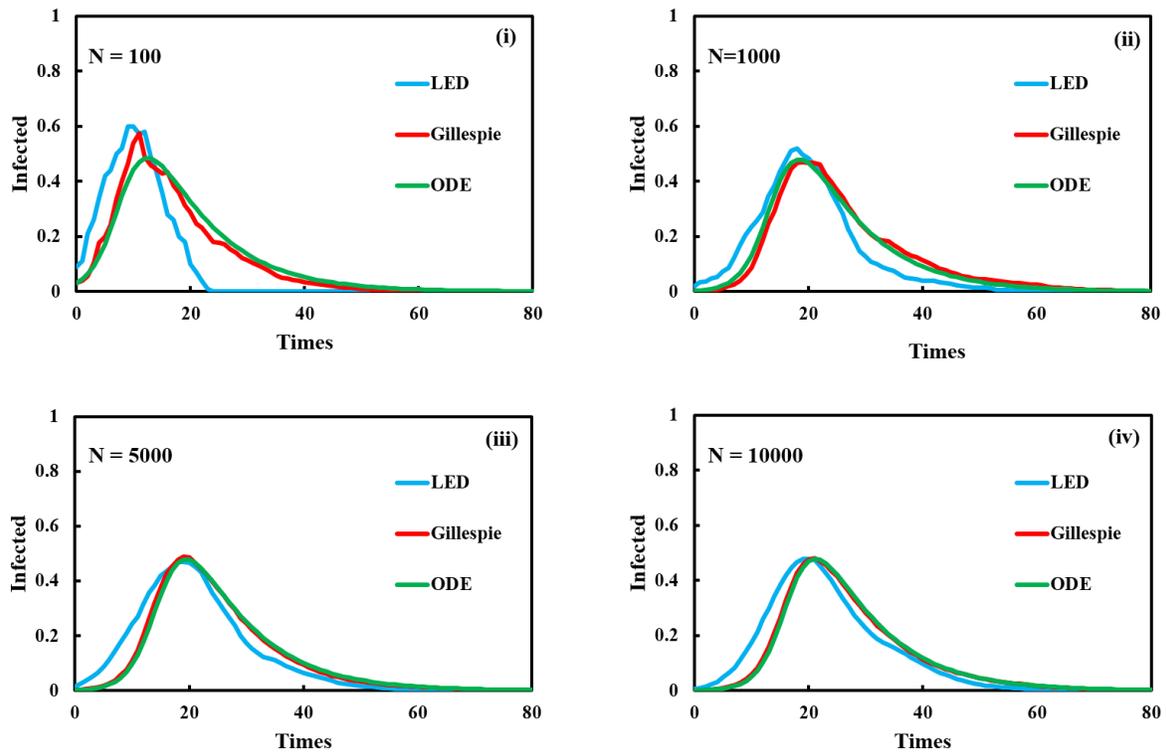

**Figure 4.** The time evolution of infected individuals is shown for varying total population sizes (N = 100, 1000, 5000, 10,000). In subfigures (i), (ii), (iii), and (iv), the results from three different models are compared: the LED model (blue), the Gillespie model (red), and the ODE model (green). These models are superimposed on each graph to highlight their consistency and demonstrate the close alignment of our proposed LED model with the established Gillespie and ODE approaches.

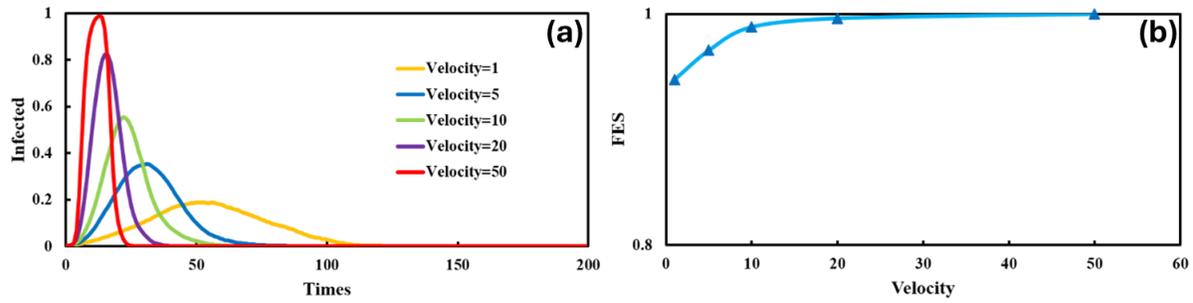

**Panel (i)**

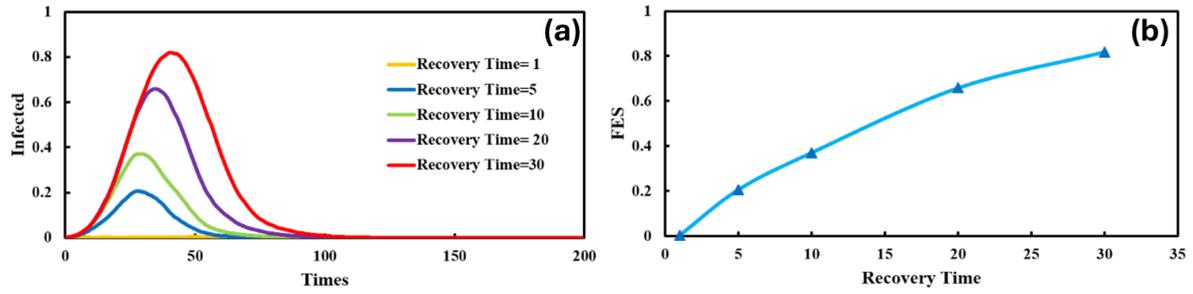

**Panel (ii)**

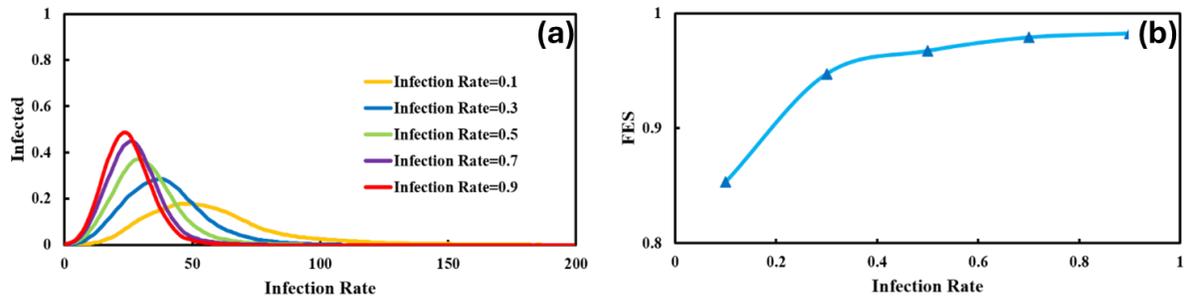

**Panel (iii)**

**Figure 5.** Epidemic dynamics and Final Epidemic Size (FES) for varying parameters without the information layer. **Panel (i)** (a) Time series of infected individuals for velocity parameters (1, 5, 10, 20, 50), colored yellow, blue, green, violet, and red. (b) FES variation as a function of velocity. **Panel (ii)** (a) Time series of infected individuals for recovery time values (1, 5, 10, 20, 30), with respective colors. (b) FES plotted against recovery time. **Panel (iii)** (a) Time series of infected individuals for infection rate values (0.1, 0.3, 0.5, 0.7, 0.9), with respective colors. (b) FES variation as a function of infection rate.

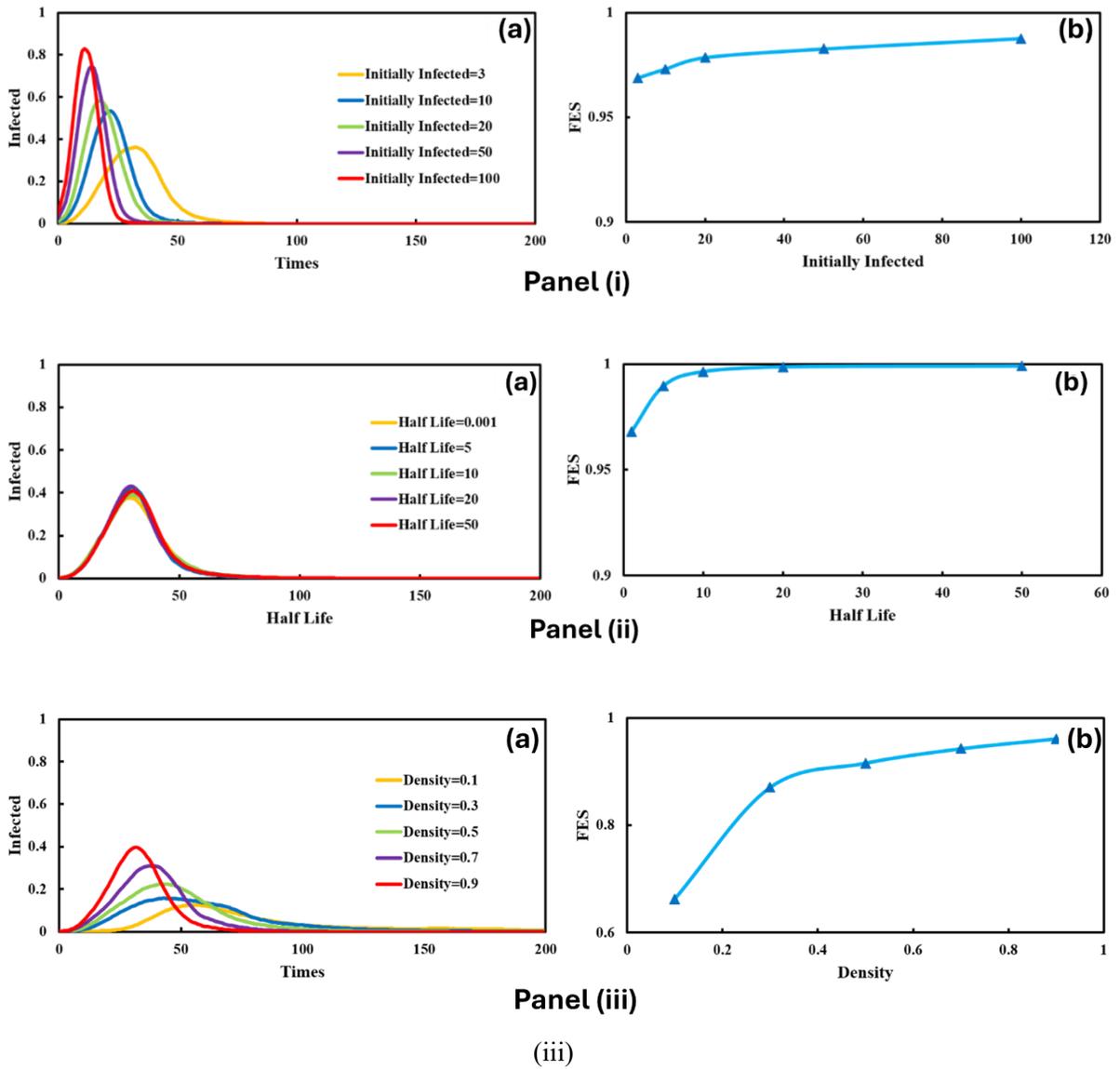

(iii)

**Figure 6.** Epidemic dynamics and Final Epidemic Size (FES) under varying parameters without the information layer. **Panel (i)** (a) Time series of infected individuals for different initial infection levels (3, 10, 20, 50, 100), colored yellow, blue, green, violet, and red, respectively. (b) FES variation as a function of initial infection levels. **Panel (ii)** (a) Time series of infected individuals for varying half-life values (0.001, 5, 10, 20, 50), with corresponding colors. (b) FES plotted against half-life values. **Panel (iii)** (a) Time series of infected individuals for varying population density (0.1, 0.3, 0.5, 0.7, 0.9), with respective colors. (b) FES as a function of population density.

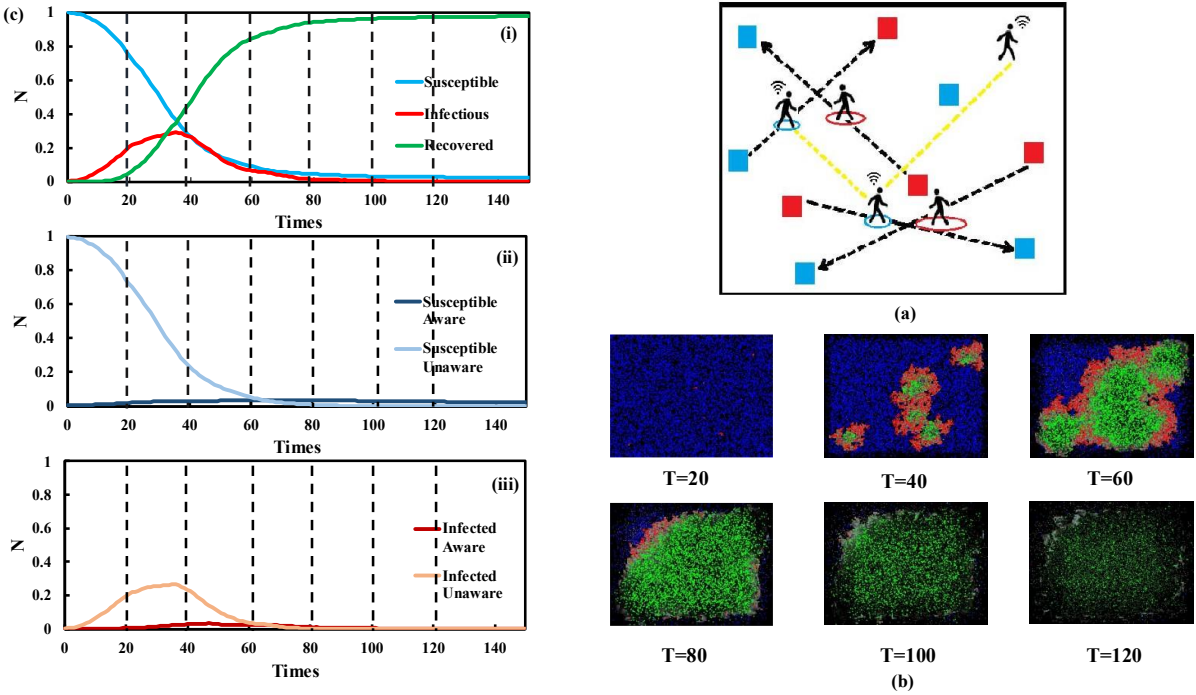

**Figure 7.** Dynamics of Susceptible, Infected, and Recovered individuals with an active information layer and mobile agents. (a) Agents move between home and office, maintaining virtual connections (yellow lines) and allowing them to share disease awareness. Aware agents (blue circles) have smaller contact radii than unaware ones (red circles), reducing their infection risk. (b) Visual representation of disease spread over time, where susceptible, infected, and recovered individuals are colored blue, red, and green, respectively. The gray background represents the 'Space' layer, indicating indirect transmission through environmental exposure. (c) The first graph shows the total number of susceptible, infected, and recovered individuals over time, including both aware and unaware agents. The second and third graphs compare infected and susceptible populations with and without the information layer, highlighting the reduced infection risk among aware agents.

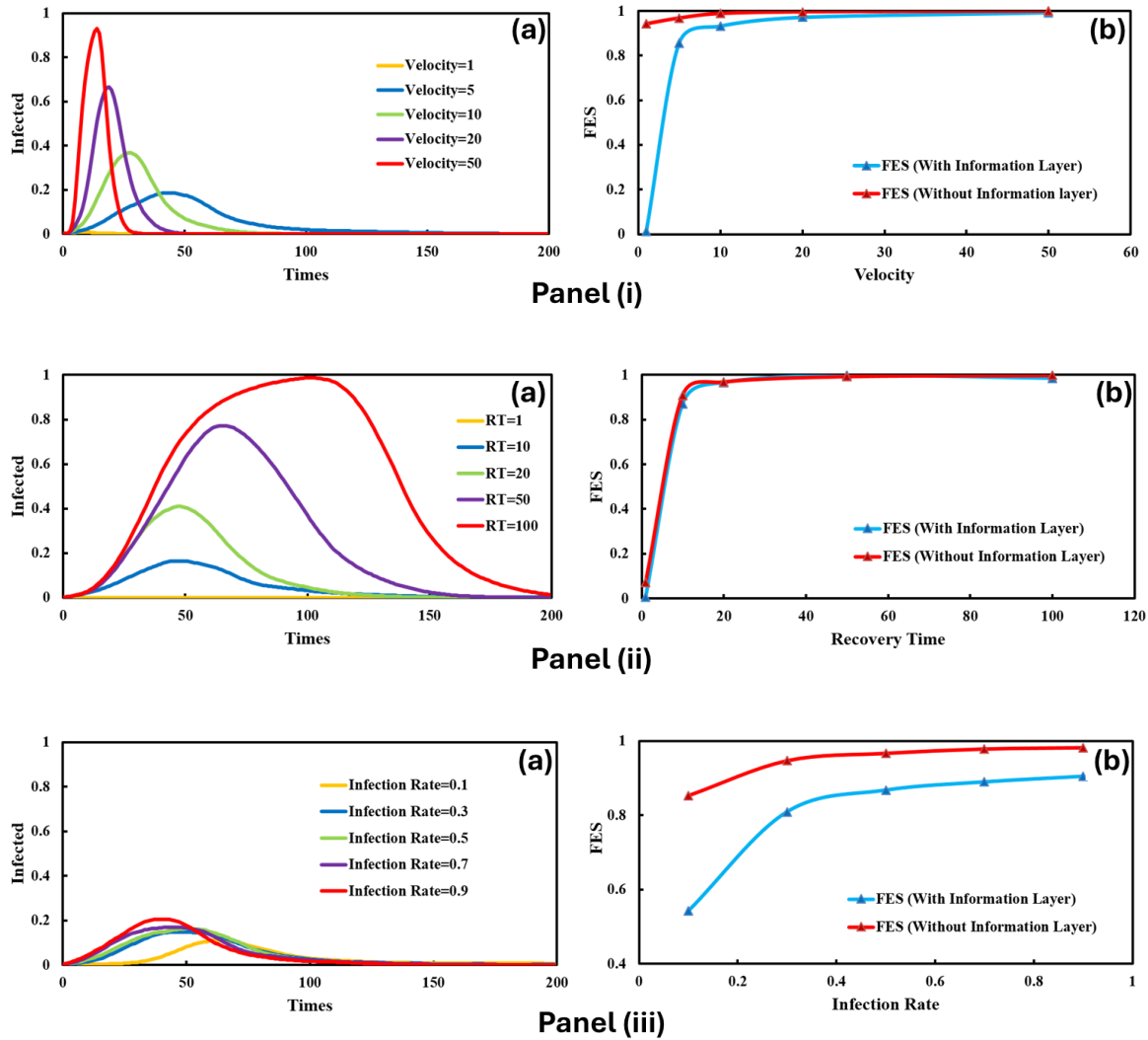

**Figure 8.** Comparison between the dynamics of epidemics and the ultimate size of the epidemic (FES) when parameters vary, both with and without an information layer. **Panel (i)** (a) Time series of infected individuals for different velocity parameters (1, 10, 20, 50, 100), with information layer colored yellow, blue, green, violet, and red, respectively. (b) FES as a function of velocity parameter, comparing scenarios with and without the information layer. **Panel (ii)** (a) Time series of infected individuals for varying recovery times (1, 10, 20, 50, 100), with corresponding colors for the information layer. (b) FES variation with recovery time for both cases. **Panel (iii)** (a) Time series of infected individuals for infection rate values (0.1, 0.3, 0.5, 0.7, 0.9), with information layer colored accordingly. (b) FES variation as a function of infection rate for both cases with and without the information layer.

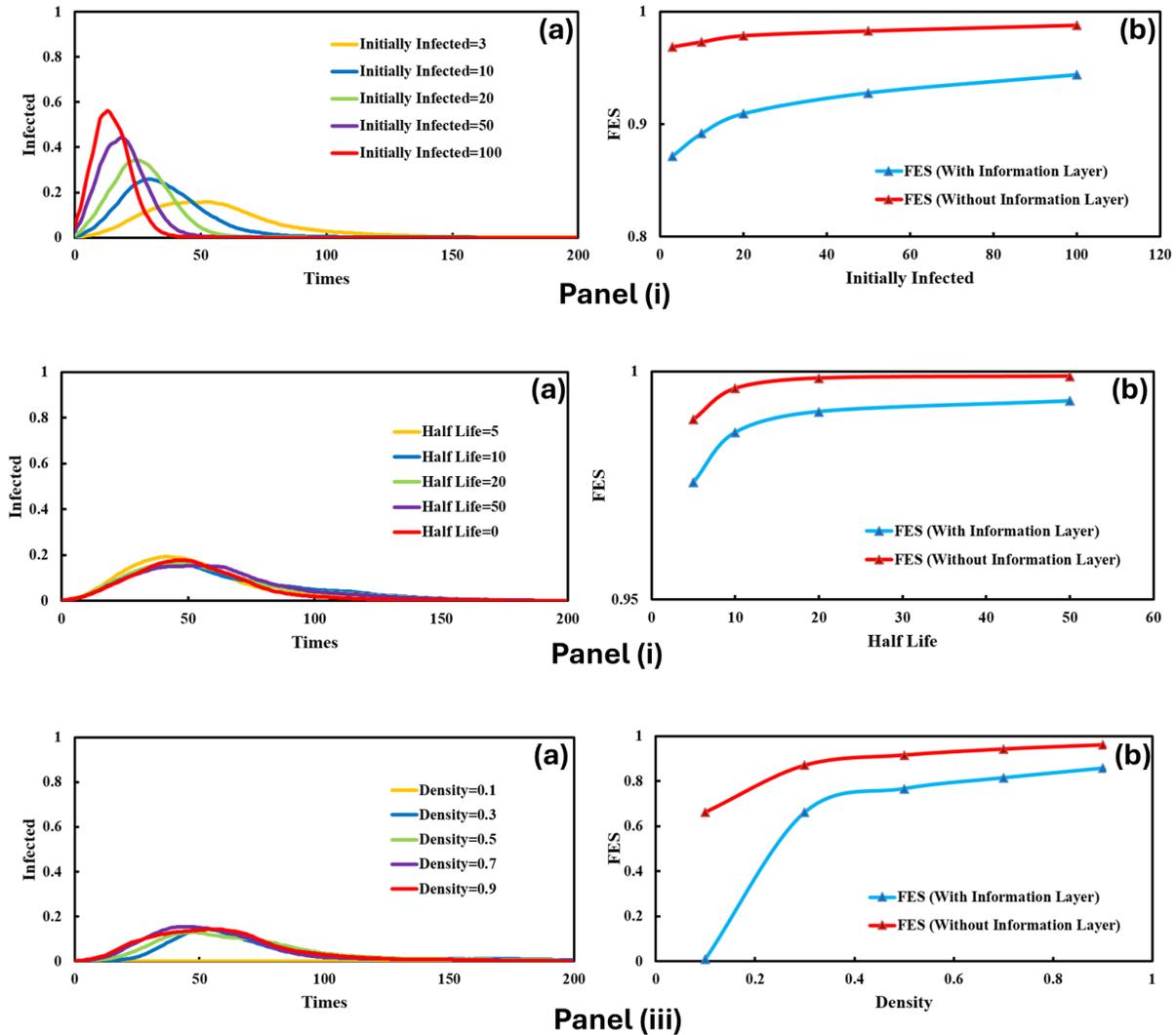

**Figure 9. Panel (i).** (a) Time evolution of infected individuals for varying initial infection levels (5, 10, 20, 50, 100) with the information layer color-coded as yellow, blue, green, violet, and red, respectively. All curves are displayed in a single graph for convenient comparison, with other parameters set to default values. (b) Final Epidemic Size (FES) plotted against the initial infection parameter, comparing cases with and without the information layer. FES values are plotted for both scenarios across all initial infection levels. **Panel (ii).** (a) Time evolution of infected individuals for varying half-life values (5, 10, 20, 50), with the information layer color-coded as yellow, blue, green, violet, and red, respectively. All curves are combined into one graph for easier comparison, with other parameters set to default. (b) Final Epidemic Size (FES) plotted against the half-life parameter, showing a comparison of FES with and without the information layer. Both cases are presented for all half-life values. **Panel (iii).** (a) Time evolution of infected individuals for varying density parameters (0.1, 0.3, 0.5, 0.7, 0.9), with the information layer color-coded as yellow, blue, green, violet, and red, respectively. The results are displayed in one graph for direct comparison, with other parameters set to default. (b) Final Epidemic Size (FES) plotted against the density parameter, comparing cases with and without the information layer. The FES values for both scenarios are plotted across all density values.